\documentclass{amsart}

\newtheorem{thm}{Theorem}[subsection]

\theoremstyle{definition}
\newtheorem{defn}[thm]{Definition}
\theoremstyle{remark}

\numberwithin{equation}{subsection}

\newcommand{\abs}[1]{\left\vert#1\right\vert}

\begin{document}

\title{Survival Strategies}

\author{ David A. Eubanks }

\address{Coker College, Hartsville, SC 29550, USA}

\email{deubanks@coker.edu}


\date{}

\dedicatory{}

\commby{}


\begin{abstract}
This paper addresses the theoretical conditions necessary for some subject of study to survive forever.  A probabilistic analysis leads to some prerequisite conditions for preserving, say, electronic data indefinitely into the future.  The general analysis would also apply to a species, a civilization, or any subject of study, as long as there is a definition of ``survival" available.  A distinction emerges between two approaches to longevity: being many or being smart.  Natural selection relies on the first method, whereas a civilization, individual, or other singular subject must rely on the latter.  A computational model of survival incorporates the idea of Kolmogorov-type complexity for both strategies to illustrate the role of data analysis and information processing that may be required.  The survival-through-intelligence strategy has problems when the subject can self-modify, which is illustrated with a link to Turing's Halting Problem.  The paper concludes with comments on the Fermi Paradox.

\end{abstract}

\maketitle

\section{Introduction}

The idea of something lasting forever is an old one.  In ancient times, the heavens were eternal and immutable. Some today speculate that human immortality is just around the technological corner \cite{singularity}. In this paper we will reduce the question of survival to probability and computation and see what this simple form can tell us about the future of life, civilizations, or just your hard drive data.

The notion of survival is more general than simply biological survival.
Whereas we might informally say that our tax records survived a hard drive
crash, we wouldn't seriously entertain the idea that the records were ever
alive in the biological sense. This generality is perhaps due to the utility of the notion of persistence in a particular state of order or function. For our purposes here, some
subject of interest ``survives" as long as it remains in some given state,
corresponding loosely to the way the word is commonly used. The only other
condition is that non-survival is permanent. For convenience, we will refer
to the transition between survival and non-survival as ``death," even though
biological life may not be involved. As an example: ``The computer died, but
the hard drive data survived." Survival in this context can be treated as a
binary condition applying to virtually any subject of interest. This could
be an individual or species of some living thing, a machine, or a more
abstract subject like a business or civilization.  In the latter realm it has interesting applications to the future of the human race, or the search for other intelligent life in the universe.

Some important related concepts are probabilities of survival, reproduction of
the subject, and the difference between the subject and its environment. In
this context, we will explore the probabilities of survival and derive
strategies for a subject to survive indefinitely long with probability greater
than zero. These strategies will illuminate the importance of reproduction
and the difficulties of surviving without it. This topic is explored further
by modeling a subject in a discrete environment where computation is assumed
to be necessary to survival in order to predict future environmental states.
This assumption allows the consideration of the computational complexity of an
environment and survival subject.

We begin with the probabilities of forever.

\section{Probabilities and Survival}

In general, we may assume that survival is uncertain and treat this as a
matter of probabilities. Imagine that some subject survives each year (or
other time period) with a probability $p$. Assuming for a moment that $p$ exists and is constant
over time, it's easy to compute the dismal odds of long term survival as a
decaying exponential. Unless $p=1$, the probability of $n$-year survival
approaches zero. The probability of surviving forever is exactly zero. But
suppose one actually wanted to survive indefinitely into the future with some
non-zero probability despite annual probabilities that are less than unity.
 Clearly, the annual probabilities must increase over time.

\subsection{Annual Survival Probabilities}

We imagine some sequence of annual conditional survival probabilities
$p_{1},p_{2,}...$ for a subject of interest. That is, the conditional
probability that our subject, having survived for $n$ years, will survive the
following year is given by $p_{n+1}$. The cumulative probability of
surviving that next year would be the product of these, so we define%

\[
P_{n}=%
{\displaystyle\prod\limits_{j=1}^{n}}
p_{j}%
\]
to be the probability of surviving $n$ years with the given probabilities, and let
$P=\lim_{n\rightarrow\infty}P_{n}$. We will be interested in cases where
$P>0$. In this case we will call the sequence $p_{1},p_{2,}...$ ``survivable."

A convenient form of $p_{n}$ for theoretical purposes is the double
exponential $p^{b^{t-1}}$, where $p\in\lbrack0,1]$ is the probability of
surviving the first year, and $b\in\lbrack0,1)$ is a change coefficient. When
time $t=1$ we see that $p_{1}=p^{b^{0}}=p$ as advertised. The second year's
conditional chance of survival, assuming the first year was survived, would be
$p_{2}=p^{b^{1}}$, and so on. Thus%
\begin{align*}
P_{n}  &  =p^{1+b+b^{2}+...+b^{n}}\\
&  =p^{(1-b^{n+1})/(1-b)}\\
&  \rightarrow p^{1/(1-b)}\text{ as }n\rightarrow\infty
\end{align*}

Let us define the ``years at risk" $y=1/(1-b)$, which can be thought of as the
(finite) number of years of risk at probability $p$ that would have to be
endured to equal the entire (infinite) number of years of risk with the
constantly increasing probability of survival controlled by $b$. For example,
if $b=.60$, then the probability of surviving forever is the same as surviving
2.5 years with a constant survival probability $p$. In this way we can
create a class of survivable probability sequences and demonstrate that there
is no mathematical objection to surviving for indefinitely long periods. If we
wish, we can compare any survivable sequence for which we know $P$ and $P_1$ to the double exponential
$p^{b^{t-1}}$ by setting $P=P_{1}^{y}$ and compute the years at risk $y=\ln
P/\ln P_{1}.$

An example of a slowly growing survivable sequence is $p_{n}=\exp
(-cn^{-1-\varepsilon})$, where $n=1,2,...$ , and $c>0$ is a scaling constant
such that $p_{1}=e^{-c}.$ We have%
\begin{align*}
P  &  =%
{\displaystyle\prod\limits_{n=1}^{\infty}}
\exp(-cn^{-1-\varepsilon})\\
&  =\exp\left(  -c%
{\displaystyle\sum\limits_{n=1}^{\infty}}
n^{-1-\varepsilon}\right) \\
&  \leq\exp\left(  -c\int_{1}^{\infty}x^{-1-\varepsilon}dx\right) \\
&  =\exp(-c/\varepsilon)\\
&  =p_{1}^{1/\varepsilon}.
\end{align*}
Because the sum in the second line is finite, $P > 0$.  The upper bound given is not particularly tight, as one can check with
$\varepsilon=1$ so that $P=\exp(-c\pi^{2}/6)=p_{1}^{\pi^{2}/6}\approx
p_{1}^{1.64}$, whereas the given upper bound is $p_{1}$. When $\varepsilon$
becomes small, $P$ also becomes small, and the years at risk are bounded by
$y\geq1/\varepsilon$. In this way we can construct survivable sequences that have arbitrarily large years at risk.

In this section we have investigated some ways in which $P>0$ can be achieved.
In all cases, there is no hope unless the annual probabilities $p_{1}%
,p_{2},..$ increase toward unity. We will next consider a practical strategy
for increasing these probabilities by making copies of the subject.

\subsection{Survival through Redundancy}

Suppose that our important data lives on a hard drive or some other medium
that has an annual probability of survival $p$. After the first year, we
copy the data onto another model of hard drive, also with probability of
survival $p$, and store it in an independent location. We continue adding
one hard drive per year ad infinitum. If any drive fails during the year, we
reconstruct it so that there are always $n$ copies at the beginning of year
$n$. Unless all the hard drives fail in a given year, our data will live
forever. For convenience, let $q=1-p$ be the chance of failure of a drive. Assuming independence of failure for the devices, we can compute the
probability of survival as%
\begin{align*}
P_{n}  &  =(1-q)(1-q^{2})...(1-q^{n}) \text{ and use a Bayesian expansion to obtain}\\
&  =1-q-q^{2}(1-q)-q^{3}(1-q)(1-q^{2})-%
{\displaystyle\sum\limits_{k=3}^{n-1}}
q^{k+1}\left\{  (1-q)...(1-q^{k})\right\}  \\
&  \geq1-q-q^{2}(1-q)-q^{3}(1-q)(1-q^{2})-(1-q)(1-q^{2})(1-q^{3})%
{\displaystyle\sum\limits_{k=3}^{n-1}}
q^{k+1}\\
&  =1-q-q^{2}(1-q)-q^{3}(1-q)(1-q^{2})-q^{4}(1-q^{2})(1-q^{3})(1-q^{n-4})\\
&  \rightarrow1-q-q^{2}(1-q)-q^{3}(1-q)(1-q^{2})-q^{4}(1-q^{2})(1-q^{3})\text{
as }n\rightarrow\infty.
\end{align*}
This inequality gives good bounds when $p$ is small. For example, when
$p=.5$ we can calculate numerically $P=.289$ versus a lower bound given by the
inequality of $P\geq.287$. The years at risk for $p=.5$ can be found to be
$y=1.79$. These calculations show that exponential growth in the number of
copies is not required for survivability--a single extra copy per unit time
suffices. However, probabilistic independence is a strong assumption. If
there is any constant probability $\varepsilon$ that the whole population will
become extinct in a year (the chance of a giant comet hitting the planet, for
example), then $P_{n}\leq(1-\varepsilon)^{n}$ still vanishes as $n\rightarrow\infty.$
This underscores the fact that independence cannot be a local condition for
physical life. To be successful, a program like the one described must
eliminate all constant probabilities of universal destruction, leaving only
ones that continue to decline.

\section{Two Strategies for Survival}

As we saw in the previous section, one has to make relatively few
independent copies in order to have a positive probability of indefinite survival, but only if survival of a copy can be considered to be general survival (unlike with an original oil painting).  In this case death doesn't occur unless all copies are destroyed.

The probability of this destruction will be lowest if the individual probabilities are uncorrelated. That is, the key to
redundancy as a strategy is that the fates of the copies are not correlated with one another. Having three backup copies of your data is great.  Having them all in the same location, not.

Life on Earth, driven by natural selection, has done a great
job of colonizing every nook and cranny of the planet, creating a very
resilient system for perpetuating life. It is apparent from these biological examples that a good strategy for increasing survival probabilities of
living things is the creation of many diverse copies. Let us agree to call a strategy of replication with variation a Multiple Independent Copies (MIC) approach. There are vast numbers of organisms on Earth (the ``multiple copies" part) with very diverse specializations (the ``independent" part).
Independence can be achieved in different ways. Physical
separation in space, differing ways of gathering energy, and tolerance to
different conditions, are some examples. No living thing on Earth is
truly independent of all others, of course, because we all inhabit the same planet. Nevertheless, we will take biological evolution as the conceptual model for a MIC.

In a MIC, survival of an individual is very much a secondary consideration to
multiplying and diversifying. It is natural to ask if it is possible for a single non-reproducing entity to prolong its life indefinitely. Let us name this singular subject a Single Intelligent System (SIS), which is intended to conjure an image of an individual organism or cohesive community that uses predictive abilities to increase conditional survival chances. ``Intelligent" refers solely to using information processing to
attempt to increase survival probabilities. This doesn't mean exactly the
same thing that that word does in common usage. A very intelligent human can
nevertheless engage in an unhealthy lifestyle. With that caveat, we'll use
the term without further disclaimers.

Lacking the massively parallel advantages of a MIC to discover and map local
environmental conditions, a SIS must do it through information processing.
It must collect information about the environment safely and inductively
predict it well enough to avoid death. Clearly, for the conditional
survival probabilities to approach one, the SIS must become better and better
at predicting the environment; it has to become smarter. In practice, a good
way to predict the environment is to learn how to control it, but we shall not
distinguish this as a separate activity.

We will contrast the MIC and SIS strategies with an abstract model of survival.
\section{A Model for the Complexity of Survival}

In this section we construct a simple computational model for a subject and its environment. The cost for this simplification is that we assume that the processes involved are computable in the sense of Church and Turing \cite{church-turing}. That is, the environment and the subject can be abstracted to a system of discrete symbols and algorithmic operations on these.

Unless an environment is completely benign, it presents both opportunities and obstacles to our subject's survival.  Taking the former and avoiding the latter must be a primary function of any subject that seeks to improve its survival changes.  Our model will allow survival only for environments where patterns of opportunities and threats can be found.  These patterns can be of arbitrary complexity.

Studying the complexity of living organisms is common, but fraught with problems in defining what complexity actually is (see \cite{biocomplexity} and \cite{growthcomplexity} for two recent example).  We will circumvent that by considering a purely theoretical model.

The complexity of patterns and rules can be defined in a useful way with the
notion of Kolmogorov Complexity, in which the complexity of a set of data is
the size of the smallest program that can reproduce it. For example, the
digits of $\pi=3.1415...$ comprise an infinite  sequence, but one that can be
compactly described with a relatively short algorithm for generating the
digits. Therefore the sequence of digits has finite complexity.  The software and hardware implementation details for this program
are generally considered to be immaterial in the study of Kolmogorov
Complexity, and the reader is referred to \cite{complexity} for more.

We will construct the following simple game, which will illuminate the relationship between
survival and environmental complexity. The two players in the game are the
subject whose survival we're investigating and the local environment in which it
finds itself. The environment generates information that the subject can
experience in the manner described below. The subject may use this
information to take action or ignore it.

\begin{defn}The Survival Game is played as follows. The environment decides on one of
three outputs: a zero, one, or blank. A
zero indicates a non-survivable environment for the subject. A one indicates a survivable condition
that would allow the subject another year (or other unit of time) of life. A blank,
denoted with a dash ``-" is a neutral
environment that neither causes the death of the subject nor aids in its
longevity. The subject must independently (without knowledge of the
environment's current play) decide on one of three actions, which are
symmetrical to the environment's. It can choose to self-destruct by playing a zero,
ending its existence. It can instead play a one, which denotes engagement
with the environment. The results of this play are either to increment
life span (in the case that the environment = 1 that turn), cause death (in the case
that the environment = 0 that turn), or have no effect (in the case that the
environment = ``-" that turn). Finally, the subject may simply ``pass" and
observe the environment by playing a blank itself, in which case it neither
increments lifespan nor dies. The subject can pass as many times as it
wishes and simply observe what the environment has delivered to it. The game
proceeds by repeating this process until the subject dies. The object of the
game is for the subject to accumulate an unbounded lifespan.
\end{defn}

The game rules are summarized in the table below, where E stands for environment, S for subject, and the result for the subject is noted in the table.

\[%
\begin{tabular}
[c]{llll}
& \textbf{E = 0} & \textbf{E = -} & \textbf{E = 1}\\
\textbf{S = 0} & Dies & Dies & Dies\\
\textbf{S = -} & No Effect & No Effect & No Effect\\
\textbf{S = 1} & Dies & No Effect & Increments Lifespan
\end{tabular}
\
\]

The elements of this game include:

\begin{itemize}
\item Information flows from environment to subject

\item Information flows from a surviving subject to environment, with a
possibility of changing it.  That is, there is the possibility that the environment's output is a function of previous plays by the subject

\item The end of the game with a finite score is compared to encountering a
non-survived condition (i.e. death)

\item It's possible for a subject to survive forever

\item The complexity of the subject and environment can vary infinitely
\end{itemize}

For any finite number of
plays we can consider the complexity of both players using the Kolmogorov
definition: complexity is the size of the smallest algorithmic description. Let us take as an example modeling the survival complexity of a fire.

A fire's response to any environmental condition is something like ``burn,
baby, burn!" That is, it doesn't pause and reflect about perhaps saving some
fuel for later or skipping that particular house out of humanitarian
concerns. So as a player in the survival game it has the lowest possible
complexity: it always plays, which we can denote \{1111...\}. The
environment may or may not be hospitable to such a strategy. In common
experience, a fire burns as long is there are suitable conditions next to it
so that it can continue to spread. A typical environment then might be some
random string of ones and zeros. Let us suppose it's \{1110...\}. It
doesn't matter what happens after the zero because the fire will be dead,
having had a lifespan of three. In this case, the environment far exceeds
the subject in complexity.

It is easy to see that a subject need not be as complex as its environment to
survive. As an example, take an environment that plays 1 on each even number
and each prime number, and 0 the rest of the time. We can imagine writing an
algorithm of fairly short length to generate this sequence by testing each odd
number for primeness. But a subject can survive the environment by simply
playing on the even numbers and passing on the odds.

It is necessary to survival that a subject anticipate and avoid all the zeros played by the environment, but it need not participate each time the environment plays one.  In order to win the game it only needs to ensure that it can always accumulate more lifespan, no matter how slowly.  All that is required is that a survivable subsequence of the environment exists, meaning that within this subsequence the subject can predict the ones and zeros, and that there are an infinite number of ones to be found. It's not necessary for the subject to be as complex as its
environment, but it needs to be as complex as some survivable subsequence.

Real environments are not static conditions that can be mastered and forgotten.  There are vastly different conditions at different locations, and they vary with time and with the actions of the subject.  We can take this into account by imagining a kind of approximation to the environmental sequence $E$ in the following sense.  Suppose that a subject masters a local condition for a while, denoted by $E_n$. As far as it knows, it has found out a survivable sequence.  To give a specific example, let's let $e_k = 1$ if $k$ is odd, and $e_k = 0$ otherwise.  This is a low-complexity rule that is understood by our subject, which flourishes in it.  But imagine that this apparent rule was only an approximation to a more complex one, and that in actuality, when $k$ is an odd multiple of 999, the environment plays zero unexpectedly, violating the conclusion of the subject's simple rule.  How can the subject negotiate this increase in complexity? We will examine that question separately from the point of view of a MIC and a SIS.

From a MIC's perspective, it's good to have lots of copies doing different things.  So it's easy to imagine that there are varieties of the subject that do not play on every single odd-numbered turn, but rather pass on some opportunities. In the scenario described above, only those specimens that pass on turn 999 will survive.  The survivors will thrive and multiply until the next die-off at turn 2997. As long as the main pattern survives with these few exceptions, and the MIC continues to make diverse copies, it can survive the higher complexity environment. The MIC as a whole has greater complexity than the simple base rule of playing each odd turn, because it can account for many different contingencies even though individuals within the MIC may have relatively low complexity.

A MIC can diversify into different local environments and develop corresponding survival strategies through trial and error, as long as it has some ability to reproduce with variation as in biological evolution.  This diversification increases the total complexity of the MIC as time goes on by virtue of the fact that it increases the number of different survivable environments. This is perhaps the simplest form of inductive reasoning: create many hypotheses and cross off the ones that don't fit the data, leaving only candidates that seem to work.  A MIC is in that sense an inductive machine for finding deductive environmental rules. As long as the environment cooperates by not changing its rules too quickly, the MIC may be able to increase in complexity indefinitely.

The author has developed a simple artificial life simulation of the survival game defined above.  It allows a user to define and change an environment, and allows a MIC of simple algorithms to evolve to survive it.  For more information see \cite{sim}.

We now turn our attention to the SIS approach to survival.  A SIS does not have the option of making copies of itself.  It can only survive by accumulating environmental ones while completely avoiding the zeros.

It is clear that a SIS needs to be cautious about participating in the environment (playing a one itself) in order to avoid an unexpected zero and death.  It must rely completely on its inductively produced prediction of environmental rules to do this.  If the environment is significantly more complex than the SIS anticipates, the SIS will miss critical contingencies and die. Imagine that we've been quietly observing the universe as it produces a subsequence  $\{101010...1010\}$ alternating ones and zeros like clockwork for some number of turns.  We are reasonably confident that we can play a one safely on the next turn of the subsequence, but we can't be certain.  This is the nature of inductive reasoning, as Karl Popper argued \cite{popper}.

In the example, the $\{101010...\}$ subsequence may eventually turn into something else, perhaps non-survivable. The only way to understand that it may end must come from outside the sequence itself.  The SIS must find some more general inductive rule that explains more of the environment.  A rule that says ``day and night alternate after several hours" cannot take into account solar eclipses.  So in order to survive, the SIS has to continually look beyond its own current survivable sequences and find patterns in other environmental data that allow broader inductive rules to be formed; it has to continually become more complex. Unlike a MIC's shotgun approach, the SIS has to do this through limited experimentation and development of theory.

A SIS would necessarily spend much of its time observing, experimenting, and formulating hypotheses in order to better refine its inductive rules (doing science, in other words).  The success of this strategy depends on probabilities for survival increasing fast enough to meet the demands outline in the first section of the paper.  Success depends on ``smoothness" of the environment, or in other words, how good induction is at predicting near-term events.

We conclude this section with some optimism that both MIC and SIS are viable strategies for long-term survival, but that MIC is more robust.  The SIS must devote most of its energy toward observation, experimentation, prediction, and careful interaction with the environment.  In the next section we will consider another challenge to long-term SIS survival.

\section{The Problem of Self-Prediction}

Any SIS would do well to control its environment in order to increase predictability. A critical observation is the following:
any SIS subject is part of its own environment; it is not
immune from creating life-threatening situations without any help from the
uncaring universe. If it must work to reduce environmental complexity to be
within its own understanding, this also applies to its own complexity. That
is, it needs to be able to model its own behavior just as much as it needs to
be able to model the behavior of external conditions. But this represents
something of a paradox. Any SIS is exactly as complex as itself. In order to
predict what it will do in the future, it actually needs to be \emph{more}
complex than itself. Understanding this challenge is the subject of this
section.

We can probably safely assume that a SIS must be able to modify
its physical parts and logical functions in order to survive indefinitely. Time and entropy will
guarantee that physical parts will have to eventually be replaced. Because
its intelligence is housed in things made of matter and energy, this means
that it can potentially modify everything about itself, both physical and
virtual.

As a thought experiment, let's imagine a well-integrated system--I'll call it
an intelligent robot. In practice, this robot could be a proxy for any
intelligent being, a whole civilization, or race of creatures integrated into
a monolithic organization. This robot is good at rationally perceiving its
environment and protecting itself from present and future external threats. But a perfectly rational being may have no particular reason for preferring existence to non-existence. In fact, a perfectly logical being may have no reason to do anything. Consider what we might call the Decider's Paradox. Our perfectly
logical robot is presented with some environmental data. What is its first
question? It could logically be ``What should my first question be?"
Similarly, its second question could logically be ``What should my second question be?"
No other types of questions are possible without an illogical answer to the
first one. Obviously some other kind of logic is desirable--something that resembles an emotional state that would allow courses of action to be prioritized.

In order for our robot to survive, it must want to, in the
sense that when it perceives threats it does not simply ignore them. Rather,
it must make complicated assessments about what trade-offs are necessary for
long-term survival, such as present sacrifice to preserve the ability to act
later. Or it might have to decide which of its appendages to sacrifice in
the present so that it may continue to function in the future. At first
blush, this seems simple enough: we'll just design our robot with this
imperative so deeply placed into its programming that it gives survival
precedence over all other things. This seems to work fine in biological organisms, which generally have a strong aversion to self death.  But for our purposes it is insufficient. Because the robot
is intelligent, it is smart enough to reverse its own engineering and perceive
this trick we've played upon it. What is to stop it from replacing that bit
of programming with something else? In fact, that may well be necessary.

Consider our own ``programming" as human beings. We're the product of
billions of years of creatures that survived long enough to reproduce, and
therefore have very deep survival instincts. And yet we can fall asleep
while driving a car. Obviously our programming didn't
foresee that possibility. If we could re-engineer that bit of our brains, we
could tweak the part that seems to say ``if not much is happening, it's okay to
sleep" and put some more sophisticated code in there. Of course we can hack
our own programming by drinking lots of coffee, opening the windows, and
tuning to a talk radio station, but this is crude compared to real self-modification. Even if we could modify our cranial code to keep us alert
while driving, it's very likely that we will not anticipate all conditions
under which we want to remain alert (or sleep). It may be
essential that we be able to self-modify in this way in order to meet future environmental conditions. But if we can do so, what is to stop us from permanently turning off the pain or hunger switch?

In terms our our survival game, let's suppose our sensible SIS has a Rule Zero that reads ``never play zero," since playing zero means death. But because it can modify its own rules, it must also
check for conditions under which it might change or eliminate Rule Zero. How might it do this? It would need to simulate all
environmental conditions to ensure that no conditions led it to become
suicidal. But this entails also tracking any changes in its own rules that
might spring from environmental changes. That is, a rule change that
resulted from energy becoming more scarce could have unexpected effects later
on, so this variant SIS would need to be simulated. But that variation would
itself spark further variations, and so on. Without a complete understanding
of what all possible rules changes and subsequent rules changes might have, it
is likely impossible to guarantee that the SIS wouldn't unexpectedly drop the Rule Zero.

If the SIS makes its decisions according to some algorithm, the situation is an example of a ``halting problem," posed by Alan
Turing \cite{halting}. The problem has no general solution in that there is no one
algorithm that can be applied to determine whether our robot will suicide or
not. That bears repeating: there is no deterministic method possible for
determining whether or not general self-modifying robots are ultimately suicidal. Actually, the situation is even worse.  A generalization called Rice's Theorem guarantees that only trivial properties of such computations can be known in general \cite{rice}.

 We might
wonder how probable this suicidal halting event is. Surprisingly, that question has
been considered by mathematicians in another form. Gregory Chaitin constructs a real number $\Omega$ that is a kind of average halting probability for algorithms, where the informal definition is
\begin{equation}
    \Omega = \sum\limits_{p \text{ halts}}{2^{\abs{-p}}}.
\end{equation}

The sum is over all programs that halt, which are expressed in some given computer language and considered binary strings for purposes of computing the sum \cite{limits}.  His
conclusion is that this number cannot be computed, nor even usefully
estimated. It inhabits a region of the real numbers that is off limits to
any algorithmic form of investigation, including conclusions based on
axiomatic systems. The theoretical interest in this number is just this
property--its pathological intractability. The $\Omega$ probability applies to a SIS only if its intelligence derives from computable behavior, a prerequisite assumption for the survival game described in the previous section.

We should therefore apply the survival probability
factor $(1-\Omega)$ to the chain of probabilities that any SIS must endure.
We cannot know what the number is, however, so we might resign ourselves to
this ``complexity tax" on the probability of survival of any SIS.
The value of $\Omega$ will not prevent an arbitrarily long-lived SIS from
existing, but it does lower the probability to some unknown upper bound.  If this upper bound is low enough, we should not expect to see a long-lived SIS in our galactic neighborhood, nor should we realistically hope that we will create one with our human endeavors.

Rice's Theorem is a general one that does not rule out the possibility of predicting what a particular algorithm will do.  There is an ongoing project, for example, to create practical tools for particular types of software projects to ensure that the programs don't hang unexpectedly.  Byron Cook, the project leader, is quoted saying:

\begin{quote}
"Turing proved that, in general, proving program termination is 'undecidable.' However, this result does not preclude the existence of future program-termination proof tools that work 99.9 percent of the time on programs written by humans. This is the sort of tool that we're aiming to make." \cite{terminator}
\end{quote}

A SIS should be heavily involved in such research activities in order to minimize the probability of self-destruction.  In addition to theoretical and experimental explorations of the environment, it must pursue similar strategies to understand its own behavior and become better at predicting it.

 \section{Conclusions}
 We have introduced the idea of analyzing the idea of indefinitely long survival with simple tools from mathematics and computer science.  Mathematically, the success of such a project depends on increasing conditional probabilities at a sufficient rate.  We have see that this is feasible using multiple independent copies (MIC) if we are allowed to consider the survival of any one copy a success. It is more challenging for a single intelligent system (SIS), which must continually improve its probabilities by developing a better understanding of its environment. These differences are usefully illustrated using a computational model for the interactions between a subject and its environment. An additional problem for a SIS is that of self-prediction in order to avoid self-destruction.

The barriers for a SIS to survive long-term are substantial. One might say that complacency is the kiss of death, since continually increasing conditional probabilities is a requirement. Given the connection to the Halting Problem, we should also conclude that careful scrutiny of the ways a SIS can self-modify must be more than a passing concern. Because of the comparative advantages of a MIC, we should not be surprised to notice that the patterns of the enduring structures of the world, both living and inert, are many and diverse rather than singular and intelligent.

These considerations lead naturally to the conclusion that if a civilization is to survive for a long time, it must become continually concerned with threats to overall survival, including internal ones.  The search for similar civilizations in the wider universe should be tempered with the likelihood that a true SIS is likely to be rare, especially in comparison to MIC-like systems.
This speaks to the Fermi Paradox, which asks why the galaxy isn't crawling with intelligent life.  A weaker form wonders why there isn't at least evidence of self-replicating ``von Neumann" probes, which could presumably spread throughout the galaxy in a few million years.

If it requires a SIS for the design of such probes, and there is a very low probability that a SIS will live long enough to develop this technology, then such a galactic swarm is less likely.  Additionally, any SIS that persists long enough to develop such technology will be able to evaluate the merits of such a plan in terms of usefulness to its own survival, and may decide that it is not in its own interests to unleash such creatures on other solar systems. This decision hinges on an interesting design problem: how adaptable should a von Neumann probe be?  Since they would have to construct copies of themselves out of raw materials they find, they would in principle have the means to completely alter their own construction.  Any errors in this process could lead to an ecology that would evolve and grow like a MIC, and this may happen before they develop the ability to colonize yet another star. Worse, they could come back and compete with the originating SIS. On the other hand, if the tendency for mutation is somehow engineered out of them, they will not be adaptable to unexpected conditions.  It may simply be that there is no `sweet spot' between these two alternatives that allows for a suitable interstellar propagation of such probes. Therefore, there may be a low probability that a SIS will create such a probe to begin with, and when they are created it may be that they evolve into local ecologies so quickly that no exponential growth of colonized systems materializes.

As a more prosaic example, the conclusions of this paper could lead one to believe that a democratic government cannot focus solely on external threats, but should also be constantly trying to improve the chances that it does not ``self-destruct" into tyranny.

As a final practical note, we might take away from this discussion the notion that if we want to increase our chances of not having an automobile accident then we should strive to be a bit more careful (or drive less) each day .  More generally, if we want the things we care about in our lives to endure, we should take a similar approach to constant improvement. Probably nothing really does last forever, but finite need not mean small.

\bibliographystyle{amsplain}
\bibliography{xbib}
\end{document}